# The Algorithm of Accumulated Mutual Influence of The Vertices in Semantic Networks

**Oleh O. Dmytrenko, Dmitry V. Lande**

*In this article the algorithm for calculating a mutual influence of the vertices in cognitive maps is introduced. It has been shown, that in the proposed algorithm, there is no problem in comparing with a widely used method – the impulse method, as the proposed algorithm always gives a result regardless of whether impulse process, which corresponds to the weighted directed graph, is a stable or not. Also the result of calculation according to the proposed algorithm does not depend on the initial impulse, and vice versa the initial values of the weights of the vertices influence on the result of calculation. Unlike the impulse method, the proposed algorithm for calculating a mutual influence of the vertices does not violate the scale invariance after increasing the elements of the adjacent matrix, which corresponds to the cognitive map, in the same value. Also in this article the advantages of the proposed algorithm on numerous examples of analysis of cognitive maps are presented.*

**Purpose:** to overcome disadvantages of the impulse method by means of a proposed in this article new algorithm for calculating mutual influence of the vertices in cognitive maps.

## 1. Introduction

A semantic network is a network that represents semantic relations between concepts. This is often used as a form of knowledge representation. It is a directed or undirected graph consisting of vertices, which represent concepts, and edges, which represent semantic relations between concepts [1].

A semantic network is used when one has knowledge that is best understood as a set of concepts that are related to one another.

Most semantic networks are cognitively based. Cognitive mapping is one of a new direction in modern decision theory [2].

A cognitive map is a directed graph, the edges of which (sometimes the vertices) have the weight. Cognitive map, like any graph, is defined by the adjacency matrix $W$ that comprises elements $w_{ij}$ – the weights of the edges, which connects to corresponding vertices $u_1, u_2, ..., u_n$. Some concepts corresponding to the vertices of a cognitive map and cause-effect (casual) relations between concepts correspond to the edges (relations).

Introduction of the weights characterizing the force of influence reveals the main focus of the cognitive approach development for the situations analysis. For well-structured situations with quantitate parameters the weights are used, and the value of influence for various paths are summarized. For fuzzy cognitive mapping analysis is well known approach, which was proposed by B. Kosko [3,4], who also introduced the term FCM (FCM – fuzzy cognitive maps). Depending on a concrete type of tasks that are solved, different modifications of FCM are considered. The basis of FCM analysis methods is the operations of fuzzy mathematics [5].

In Kosko model an influence is calculated by the following way. An indirect influence action (An indirect effect) $I_p$ of vertex $i$ on vertex $j$ through the path $P$ that directs from vertex $i$ to vertex $j$ is defined as

$$I_p = \min_{(k,l) \in E(P)} w_{kl},$$

where $E(P)$ is a set of edges of the path $P$; $w_{kl}$ is weight of the edge $(k,l)$ of the path $P$, which value is defined by the terms of linguistic variables.

The general influence $T_{(i,j)}$ of vertex $i$ on vertex $j$ is defined as

$$T_{(i,j)} = \max_{P(i,j)} I_p,$$

where the maximum is taken from all ways that follow from vertex $i$ to vertex $j$.

As a result, $I_p$ defines the weakest link of the path $P$, and $T_{(i,j)}$ defines the strongest influence among the indirect influences $I_p$.

Also for fuzzy cognitive mapping analysis the Delphi Method is used.

For example, in the work [6] the Delphi Method is used for determining the relationships between factors, which effect the planning process on the purpose of building strategic information systems.

## Impulse method

One of the methods of cognitive mapping analysis is the impulse method, which was introduced in $1970^{th}$ [7,8]. According to the impulse method, each vertex takes a value $v_i(t)$ in a discrete moment of time $t = 0,1,2,...$. The weight of the edge has a positive value ($w_{ij} > 0$) if increasing the weight of the vertex $u_i$ leads to the increasing of the weight of the vertex $u_j$. And vice versa, the weight of the edge has a negative value ($w_{ij} < 0$) if decreasing of the weight of the vertex $u_i$ leads to decreasing of the weight of the vertex $u_j$. $w_{ij} = 0$ if the vertices $u_i$ and $u_j$ are not related.

The value $v_i(t)$ of each vertices changes in a discrete moment of time $t = 0,1,2,...$ according to the equation:

$$v_i(t+1) = v_i(t) + \sum_{j=1}^{n} w(u_i, u_j) p_j(t) \qquad (1)$$

where $n$ is an amount of the vertices in the graph.

The impulse is defined by the following equation:
$$p_j(t) = v_i(t) - v_i(t-1), \ t > 0.$$

At the initial moment $t = 0$: $p_j(0)$ and $v_i(0)$ are defined.

## Disadvantages of impulse method

Despite widespread use of the impulse method, it has some disadvantages.

The main and most important disadvantage of the impulse method is a diverging of the calculation result in case when an impulse process, which corresponds to the weighted directed graph, is not stable.

If all non-zero characteristic constants of the weighted directed graph defined by the adjacency matrix $W$ are different and absolute value does not exceed the one, then the directed graph is impulse stable for all simple impulse processes. Otherwise, the weighted directed graph is unstable within the meaning of impulse for some simple impulse process [7]. This means that the vertex will be found, which gets the initial impulse. The impulse in some (maybe another) vertex will become infinitely large. In other words, in the impulse method the value $v_i(t)$ is not defined when $t \to \infty$. That is why methods of stabilization exist [9]. Also, in [10] a critical analysis of the main methods of cognitive maps researching has been carried out and a number of drawbacks and contradictions that arise in applying the impulse method are given:

1) Diverging $v_i(t)$ with $t \to \infty$, in $\infty$ step, when series in (1) diverge.

2) The result of calculation - $v_i(t)$ depends on according to (1) on the initial values to $p_j(0)$.

3) Initial value $v_i(0)$ does not at all affect the dependence $v_i(t)$ from $t$ (comes into the expression for $v_i(t)$ as item).

4) Increasing the values of each elements of matrix $W$ in the same value, does not only change the value component of vector $v(t)$, but changes their rank of distribution. In some cases it leads to the numeric series diverging (1).

## The algorithm for calculating mutual influence of vertices

In this article a new approach for cognitive maps investigation – the algorithm for calculating a mutual influence of the vertices in cognitive maps is proposed. The idea of this algorithm is that all vertices of a weighted directed graph are considered in pairs and the value of influence $z_{ij}$ of vertex $u_i$ on vertex $u_j$ is calculated (where $i, j = 1, 2, ..., n$). As a result the influence matrix $Z$ consisting of elements $z_{ij}$ is obtained. For calculating the value of general influence $z_{ij}$ of vertex $u_i$ on vertex $u_j$ the next steps has been taken:

1) all possible simple ways from the vertex $u_i$ to the vertex $u_j$ are built. The block

diagram of a recursive algorithm used is presented on the fig.1

At the input the pair of the vertices $(u_i, u_j)$ are given, where $u_i$ is a vertex the impact of which is determined (the initial vertex) and $u_j$ is the vertex under the influence (the final vertex).

I. A step in the opposite direction to the edge from the final vertex $u_j$ into the initial vertex $u_i$ is taken. The vertex $u_j$ is added to the path. The last added vertex is the current.

II. "Conditional test 1" means that the verification is performed.
"Conditional test 1" means the verification of whether a vertex directly reachable from the current vertex (path of length 1) but not included into the path exists. If it is true then "Conditional test 2" is made.
Otherwise, the path that is built is a deadlock, and the last added vertex is deleted from the path.

III. At the stage "Conditional test 2" a check whether a vertex directly reachable from the current vertex (path of length 1) coinciding with the initial vertex $u_i$ is performed. If it does not coincide then the vertex is added to the path and the next step is performed from this vertex (the vertex becomes the current). The next is step II.
Otherwise, if the vertex directly reachable from the current vertex (path of length 1) coincides with the initial vertex $u_i$ then this vertex is added to the path. The found path is displayed on the screen in a reverse order. The next, the previous added vertex is deleted from the path and a move to step II is performed.

2) at every path the influence from the vertex $u_i$ on the vertex $u_j$ is calculated, considering the weights of the edges: the impulse from the vertex $u_i$ is distributed in the chain from $u_i$ to $u_j$ according to the rules a) – d) [11]:

a) $u_i \xrightarrow{+} u_k \xrightarrow{-} u_j$

If the vertex $u_i$ has a positive influence on the vertex $u_k$, and $u_k$ has a negative influence on $u_j$, then the vertex $u_i$ increases the negative influence of $u_k$ on $u_j$. As a result the vertex $u_i$ has a positive influence on $u_j$.

b) $u_i \xrightarrow{-} u_k \xrightarrow{-} u_j$

If the vertex $u_i$ decreases a negative influence of the vertex $u_k$ on $u_j$, then the vertex $u_i$ has a positive influence on $u_j$.

c) $u_i \xrightarrow{+} u_k \xrightarrow{+} u_j$

In this case the $u_i$ has a positive influence on $u_j$, increasing a positive influence of the vertex $u_k$ on $u_j$.

d) $u_i \xrightarrow{-} u_k \xrightarrow{+} u_j$

In this case, the vertex $u_i$ has a negative influence on the vertex and $u_k$ has a positive influence on $u_j$, then the vertex $u_i$ decreases the positive influence of $u_k$ on $u_j$ As a result the vertex $u_i$ has a negative influence on $u_j$.

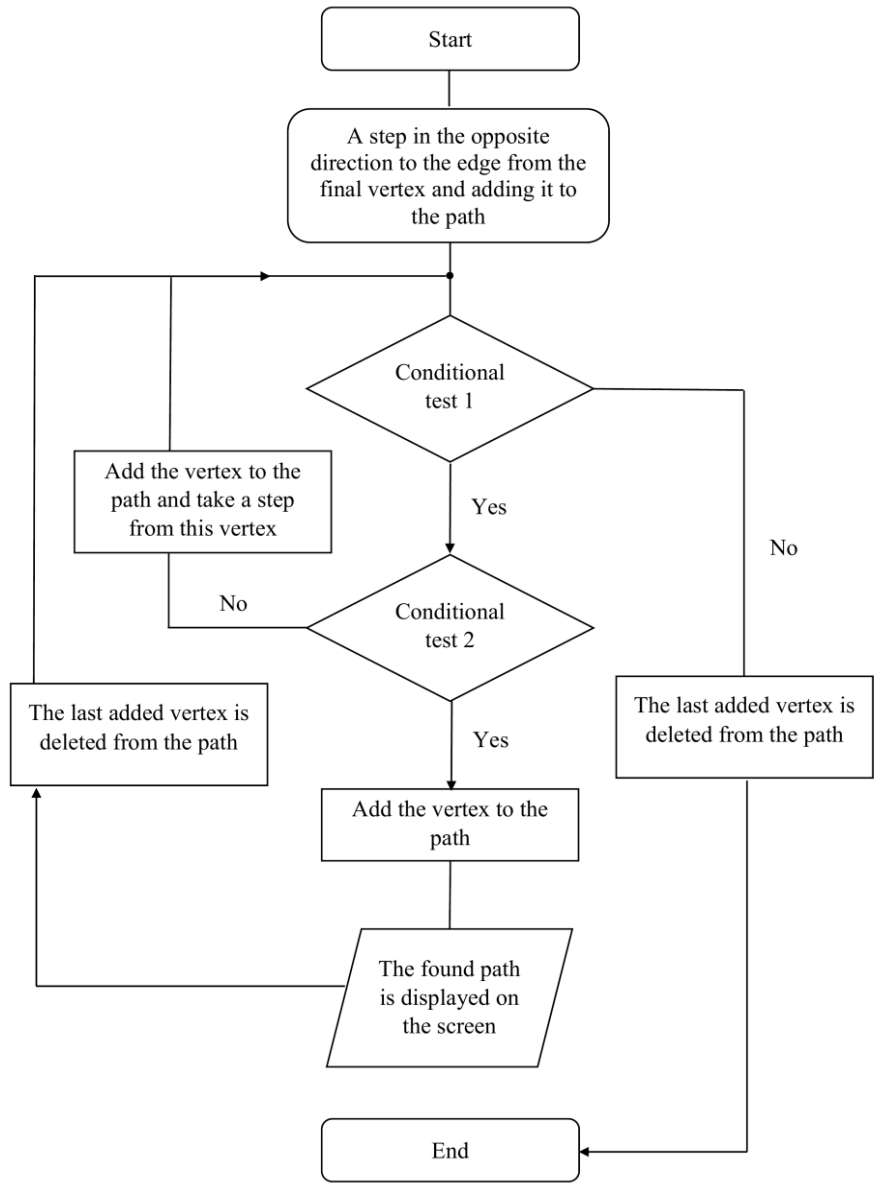

Fig. 1 Flowchart of an algorithm for building all possible simple ways between a pair of vertices of the cognitive map

For calculating the partial influence on the final vertex $u_j$ accumulated form the vertex $u_i$ on the $k$ simple path, it is necessary to calculate the general influence $z_{ij}^k$ on the vertex $u_j$, which is accumulated from all vertices $q_t^k$ that are involved in the $k$-way (taking into account the rules a)–d) ); then to subtract from $z_{ij}^k$ the influence $\tilde{z}_{ij}^k$ on the vertex $u_j$, which is accumulated from all vertices $q_t^k$ involved into the $k$-way, which does not involve the initial vertex $q_0 = u_i$.

$z_{ij}^k$ and $\tilde{z}_{ij}^k$ are calculated iteratively according to the formulas:

$$z_{ij}^k(t+1) = \left(1 + \text{sign}(z_{ij}^k(t)) * \alpha\left(\left|\frac{z_{ij}^k(t)}{\mu}\right|\right)\right) * w(q_t^k, q_{t+1}^k) \quad (2)$$

$$\tilde{z}_{ij}^k(r+1) = \left(1 + \text{sign}(\tilde{z}_{ij}^k(r)) * \alpha\left(\left|\frac{\tilde{z}_{ij}^k(r)}{\mu}\right|\right)\right) * w(q_r^k, q_{r+1}^k) \quad (3)$$

where $q_t^k$ is a sequence of the vertices involved in a $k$-way ($q_0 = u_i, q_{m-1} = u_j$); $t = 0,1,...,m-2$, and $r = 1,...,m-2$ (where $m$ is a number of vertices that are involved in the $k$-way).

The initial conditions:

$z_{ij}^k(0) = 0$, $\tilde{z}_{ij}^k(1) = 0$,

$\mu = \max|w_{ij}|$,

where $i = 0,1,...,n$, $j = 0,1,...,n$ ($n$ is a dimension of the cognitive map).

The general influence $z_{ij}$ on the vertex $u_j$ accumulated from the vertex $u_i$ is the sum of the partial influences calculated as the delta between the (2) and the (3) on all simple ways from the vertex $u_i$ into the vertex $u_j$

$$z_{ij} = \sum_{k=1}^{s}(z_{ij}^k - \tilde{z}_{ij}^k)$$

where $s$ is a number of all simple ways from the vertex $u_i$ to the vertex $u_j$.

If the vertex $u_j$ is not reachable from the vertex $u_i$, then it means that $z_{ij} = 0$.

The $\mu$-normalizing on every step makes opportunities to avoid a large value of the coefficient $\alpha$ and adequately display the influence of a previous vertex.

The coefficient $\alpha(x) = 1 - e^{-2x}$, $x \geq 0$ is the function of a stochastic distribution for an exponential distribution with the parametric variable that equals to $\lambda = 2$.

The proposed algorithm has an exponential complexity $O(e^{2n})$, where $n$ is a number of vertices of a cognitive map. It means that after the increase of a number of the cognitive map vertices, a number of iterations will increase exponentially, that will lead to the increase of a time complexity. A process of building of all simple ways between two vertices has a great contribution in a computational complexity, because in each of the vertices it is necessary to choice the next vertex not added to path.

Also for decreasing a number of iterations (as consequently for decreasing of the main load on computing complexity), the initial step is proposed, which main idea is:

a) to build the reachable matrix $A$ for the initial matrix $W$;

b) every element $w_{ij}$ of the initial matrix $W$ to multiply on the corresponding

element $a_{ij}$ of the matrix $A$.

A new obtained sparse matrix $W$ gives an opportunity to find beforehand all possible simple ways between vertex $u_i$ and vertex $u_j$ (step 1) to define whether the vertex $u_j$ is reachable from the vertex $u_i$.

Accordingly if the way from vertex $u_i$ to vertex $u_j$ does not exist (in a sparse matrix $w_{ij}=0$), then the element $z_{ij}$ of a matrix $Z$ is equal to zero ($z_{ij}=0$), which means that the vertex $u_i$ does not influence the vertex $u_j$.

### Advantages and examples of the algorithm work for calculating mutual influence of vertices

One of the advantages of the proposed algorithm, unlike the impulse method, is the limitedness of the algorithm result $z_{i,j}$ for any finite number $n$ of vertices and for any values $w_{ij}$ of the weighted directed graph.

For $\forall x \geq 0$ the coefficient $\alpha(x)$, which characterizes the accumulative influence, is limited $0 \leq \alpha(x) < 1$.

As the result of any finite values $r$ and $t$ of the value $1 \pm \alpha(x)$ is limited $0 < 1 \pm \alpha(x) < 2$, the values $z_{ij}^k(r)$ and $\tilde{z}_{ij}^k(t)$ corresponding to (2) and (3) are limited: accordingly $0 < |z_{ij}^k(r)| < 2*\mu$ and $0 < |\tilde{z}_{ij}^k(t)| < 2*\mu$. Then the difference $z_{ij}^k - \tilde{z}_{ij}^k$ that characterizes the value of influence of vertex $u_i$ on vertex $u_j$ for a $k$-way, is also limited: $-2*\mu < z_{ij}^k - \tilde{z}_{ij}^k < 2*\mu$. As a result the sum $\sum_{k=1}^{s}(z_{ij}^k - \tilde{z}_{ij}^k)$ is limited $-2*\mu*s < \sum_{k=1}^{s}(z_{ij}^k - \tilde{z}_{ij}^k) < 2*\mu*s$ for $\forall s > 0$ (where $s$ is a number of all simple ways from the vertex $u_i$ to the vertex $u_j$). Since $s$ and $\mu$ are finite, then $\sum_{k=1}^{s}(z_{ij}^k - \tilde{z}_{ij}^k)$ is limited $\sum_{k=1}^{s}(z_{ij}^k - \tilde{z}_{ij}^k)$ because $s$ and $\mu$ are limited. It means that for any finite number $n$ of vertices as a result of limitedness (because of the $\alpha(x)$ is limited), the result $z_{i,j}$ exists and limited by a number value. For example, for complete graph with a dimension $n$, a number of a simple ways $s$ from the vertex $u_i$ to the vertex $u_j$ is:

$$s = \left(1 + \frac{1}{2!} + \frac{1}{3!} + ... + \frac{1}{(n-3)!} + \frac{1}{(n-2)!}\right) * (n-2)! < e * (n-2)!.$$

In this case the result of calculating $-z_{i,j}$ is limited:

$$-2*\mu*e*(n-2)! < z_{i,j} < 2*\mu*e*(n-2)!$$

The graph of a simple impulse process defined by the adjacent matrix:

$$W = \begin{pmatrix} 0 & 0{,}391 & -0{,}121 & 0 \\ 0 & 0 & 0 & 1 \\ 0 & 0 & 0 & -1 \\ 1 & 0 & 0 & 0 \end{pmatrix} \quad (4)$$

and corresponded to the weighted directed graph (fig. 2) is shoved on the fig. 3.

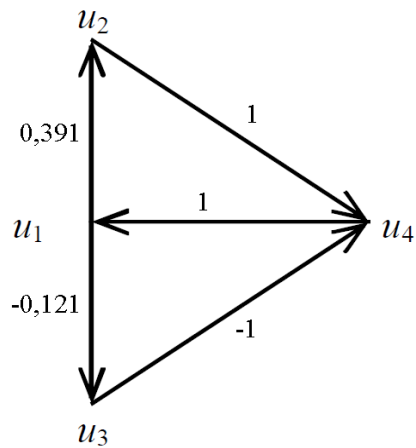

Fig. 2 Weighted directed graph

All non-zero characteristic constants of the adjacency matrix $W$, which describes the weighted directed graph, are different and equal to $0{,}8$; $0{,}4(-1+i\sqrt{3})$; $0{,}4(-1-i\sqrt{3})$, and their absolute value do not exceed the one, therefore the directed graph defined by (4) is impulse stable for all simple impulse processes. The directed graph defined by (4) is impulse stable for all simple impulse processes, therefore among all characteristic constants the value that exceeds the one does not exist. For example, if the vertex $u_1$ gets an unit impulse, on the $61^{th}$ iterative step the impulse method converge completely (fig. 3).

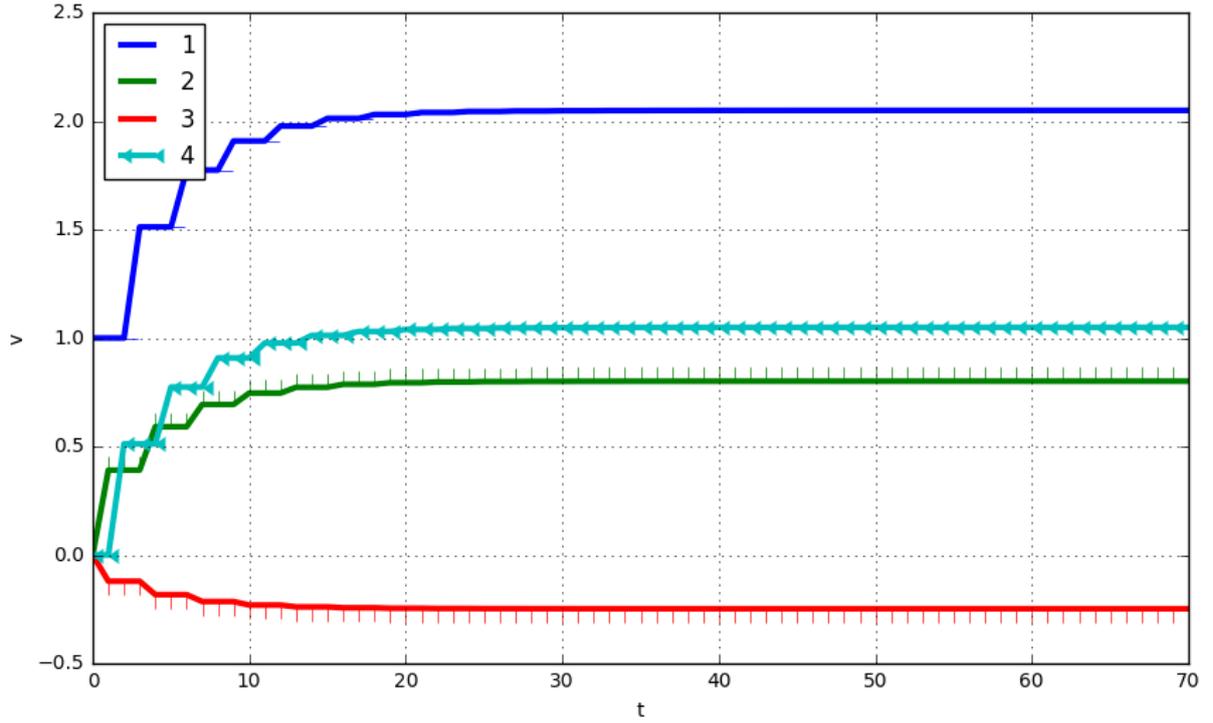

Fig. 3 Graph of a simple impulse process (the influence of the vertex $u_1$, when $u_1$ gets a unit impulse)

The general influence $Inf_{sm}$ of each vertex and their ranking for the matrix (4) are presented in the tab. 1.

Table 1

| Vertex (No) | $Inf_p$ |
|---|---|
| 3 | 4,899 |
| 2 | 4,346 |
| 4 | 3,098 |
| 1 | 2,098 |

The influence matrix obtained as a result of the algorithm use for calculating a mutual influence of the vertices is

$$Z = \begin{pmatrix} 0 & 0,391 & -0,121 & 0,757 \\ 0,865 & 0 & -0,013 & 1 \\ -0,865 & -0,245 & 0 & -1 \\ 1 & 0,338 & -0,105 & 0 \end{pmatrix} \quad (5)$$

The general influence $Inf_{am}$ of each vertices for the influence matrix $Z$ is defined according to the rule:

$$Inf_{am}^{i} = \sum_{j=1}^{n} |z_{ij}|, \quad (6)$$

where $n$ is a number of vertices.

The general influence $Inf_{am}$ of each vertex and their ranking for the matrix (5), according to (6), are presented in the tab. 2.

Table 2

| Vertex (No) | $Inf_{am}$ |
|---|---|
| 3 | 2,11 |
| 2 | 1,878 |
| 4 | 1,443 |
| 1 | 1,269 |

Tab. 1 and tab. 2 are demonstrating that the ranking of vertices for the impulse method and for the algorithm for calculating a mutual influence of the vertices does not change.

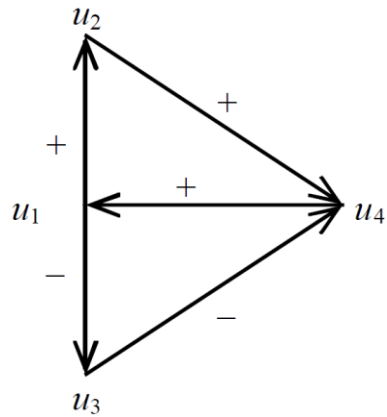

Fig. 4 Weighted directed graph for unstable impulse process

All non-zero constant characteristics of the adjacency matrix:

$$W = \begin{pmatrix} 0 & 1 & -1 & 0 \\ 0 & 0 & 0 & 1 \\ 0 & 0 & 0 & -1 \\ 1 & 0 & 0 & 0 \end{pmatrix}, \qquad (7)$$

which describes the impulse process (fig .4), are not different and exceed the one $(1,26;\ 1,26;\ 1,26)$.

Then the directed graph defined by (7) is unstable within the meaning of impulse for some simple impulse process. In another words, there is the vertex which obtains a unit impulse, that in some (maybe another) vertex, the impulse becomes an indefinitely large. In this case the impulse method is diverged and the value $v(\infty)$ is not defined. As a result, the algorithm for calculating a mutual influence of the vertices in cognitive maps is used. The influence matrix for the matrix (7) is

$$Z = \begin{pmatrix} 0 & 1 & -1 & 1{,}729 \\ 0{,}865 & 0 & -0{,}111 & 1 \\ -0{,}865 & -0{,}628 & 0 & -1 \\ 1 & 0{,}865 & -0{,}865 & 0 \end{pmatrix} \quad (8)$$

The general influence $Inf_{am}$ of each vertex and their ranking for the matrix (8) are presented in the tab. 3.

Table 3

| Vertex (No) | $Inf_{am}$ |
|---|---|
| 1 | 3,729 |
| 4 | 2,729 |
| 3 | 2,492 |
| 2 | 1,976 |

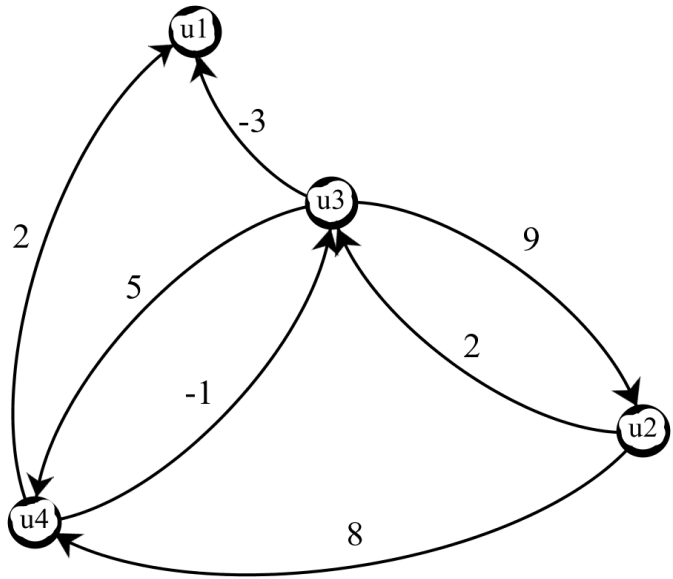

Fig. 5 Weighted directed graph

The adjacency matrix for the weighted directed graph (fig. 5) looks as

$$W = \begin{pmatrix} 0 & 0 & 0 & 0 \\ 0 & 0 & 2 & 8 \\ -3 & 9 & 0 & 5 \\ 2 & 0 & -1 & 0 \end{pmatrix} \quad (9)$$

All non-zero constant characteristics of the adjacency matrix (9), which describes the weighted directed graph (fig. 5) are equal: $5{,}92;\ 3{,}48;\ 3{,}48$.

Therefore, the directed graph defined by (9) is unstable within the meaning of impulse for some simple impulse process.

The influence matrix for the adjacency matrix (9) is

$$Z = \begin{pmatrix} 0 & 0 & 0 & 0 \\ 1,207 & 0 & 1,169 & 9,794 \\ -1,393 & 9 & 0 & 11,917 \\ 2,598 & -1,79 & -1 & 0 \end{pmatrix} \qquad (10)$$

The general influence $Inf_{am}$ of each vertex and their ranking for the matrix (10) are presented in the tab. 4.

Table 4

| Vertex (No) | $Inf_{am}$ |
|---|---|
| 3 | 22,31 |
| 2 | 12,17 |
| 4 | 5,39 |
| 1 | 0 |

In the work [12] the weighted directed graph is considered (fig. 6). This graph is built for the analysis of the problem of solid waste removal from cities (the example is taken from "Maruyama M. The Second Cybernetics: Deviation-Amplifying Mutual Causal Processed. – Amer. Scientist. – 51. – 1963. – P. 164-179").

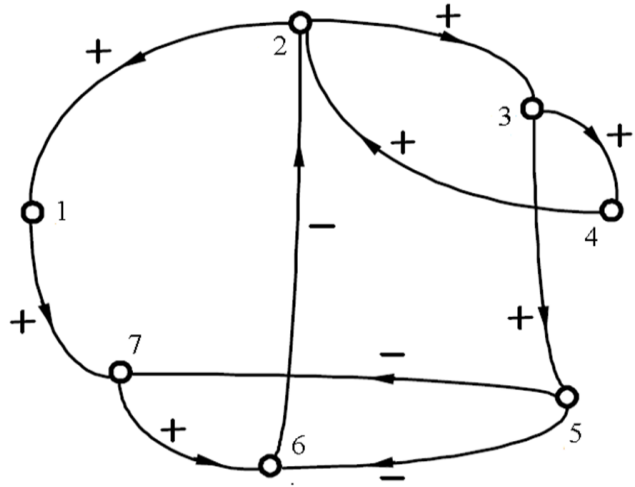

Fig. 6 Weighted directed graph of the analysis of the problem of solid waste removal from cities: 1 – Amount of garbage per unit area; 2 – Number of inhabitants in the city; 3 – Improving living conditions in the city; 4 – Migration to the city; 5 – Number of treatment facilities; 6 – Number of diseases; 7 – Bacteriological contamination per unit area

For the weighted graph presented on the fig. 6, the adjacency matrix is

$$W = \begin{pmatrix} 0 & 0 & 0 & 0 & 0 & 0 & 1 \\ 1 & 0 & 1 & 0 & 0 & 0 & 0 \\ 0 & 0 & 0 & 1 & 1 & 0 & 0 \\ 0 & 1 & 0 & 0 & 0 & 0 & 1 \\ 0 & 0 & 0 & 0 & 0 & -1 & -1 \\ 0 & -1 & 0 & 0 & 0 & 0 & 0 \\ 0 & 0 & 0 & 0 & 0 & 1 & 0 \end{pmatrix} \quad (11)$$

The directed graph defined by the matrix (11) is unstable within the meaning of impulse for some simple impulse process because absolute values of all non-zero characteristic constants of the adjacency matrix (11) are equal: $1,19$; $1,09$; $1,09$; $0,84$; $0,84$; $1,14e-16$; $4,79e-16$.

The influence matrix for the matrix (11) is

$$Z = \begin{pmatrix} 0 & -0,111 & -0,005 & -0,009 & -0,009 & 0,865 & 1 \\ 1 & 0 & 1 & 0,865 & 0,865 & -0,005 & 0,753 \\ 0,34 & 1,166 & 0 & 1 & 1 & -0,975 & -0,791 \\ 0,865 & 1 & 0,865 & 0 & 0,111 & -0,0002 & 0,107 \\ 0,688 & 1,492 & 0,688 & 0,654 & 0 & -1,865 & -0,455 \\ -0,865 & -1 & -0,865 & -0,628 & -0,628 & 0 & -0,567 \\ -0,111 & -0,864 & -0,111 & -0,19 & -0,19 & 1 & 0 \end{pmatrix} \quad (12)$$

The general influence $Inf_{am}$ of each vertex and their ranking for the matrix (12) are presented in the tab. 5.

Table 5

| Vertex (No) | $Inf_{am}$ |
|---|---|
| 5 | 5,841 |
| 3 | 5,274 |
| 6 | 4,552 |
| 2 | 4,487 |
| 4 | 2,947 |
| 7 | 2,468 |
| 1 | 1,999 |

On the fig. 7 an example of the weighted graph of a cognitive model of electricity consumption is considered (the example is taken from "*Roberts F.S.* Signed

Diagraphs and the Growing Demand for Energy, Environment and Planning. – 1971. – P. 395-410")

For the weighted graph presented on the fig. 7, the adjacency matrix is

$$W = \begin{pmatrix} 0 & -1 & 0 & 0 & 0 & 0 & 1 \\ 0 & 0 & -1 & 0 & 0 & 0 & 0 \\ 1 & -1 & 0 & -1 & 0 & 0 & 0 \\ 0 & 0 & 0 & 0 & 1 & 0 & 0 \\ 0 & 0 & 1 & 0 & 0 & 0 & 0 \\ 0 & 0 & 0 & 0 & 1 & 0 & 0 \\ 0 & 0 & 1 & 0 & 0 & 1 & 0 \end{pmatrix} \quad (13)$$

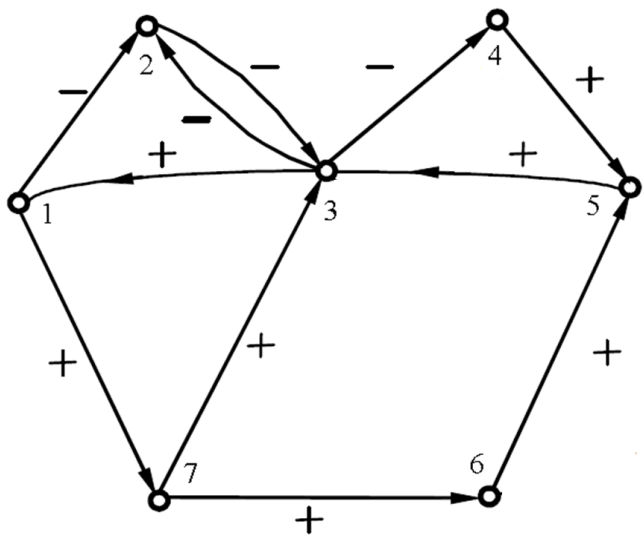

Fig. 7 Example of the weighted graph of a cognitive model of electricity consumption: 1 – Power capacity; 2 – Cost of electricity; 3 – Power consumption; 4 – State of the environment; 5 – The population; 6 – Number of workplaces; 7 – Number of enterprises

The directed graph defined by matrix (13) is unstable within the meaning of impulse for some simple impulse process because absolute values of all non-zero characteristic constants of the adjacency matrix (13) are equal: $1,429$; $1,125$; $1,125$; $0,743$; $0,743$; $9,97$.

The influence matrix for the matrix (13) is

$$Z = \begin{pmatrix} 0 & -1{,}112 & 1{,}734 & -0{,}739 & 0{,}439 & 0{,}865 & 1 \\ -0{,}865 & 0 & -1 & 0{,}865 & 0{,}625 & -0{,}06 & -0{,}628 \\ 1 & -1{,}865 & 0 & -1 & -0{,}86 & 0{,}111 & 0{,}865 \\ 0{,}111 & -0{,}116 & 0{,}865 & 0 & 1 & 0{,}0002 & 0{,}005 \\ 0{,}865 & -0{,}976 & 1 & -0{,}865 & 0 & 0{,}005 & 0{,}111 \\ 0{,}111 & -0{,}116 & 0{,}865 & -0{,}111 & 1 & 0 & 0{,}005 \\ 0{,}869 & -0{,}981 & 1{,}11 & -0{,}869 & 0{,}753 & 1 & 0 \end{pmatrix} \quad (14)$$

The general influence $Inf_{am}$ of each vertex and their ranking for the matrix (14) are presented in the tab. 6.

Table 6

| Vertex (No) | $Inf_{am}$ |
|---|---|
| 1 | 5,888 |
| 3 | 5,7 |
| 7 | 5,585 |
| 2 | 4,042 |
| 5 | 3,821 |
| 6 | 2,208 |
| 4 | 2,097 |

Also the advantage of the algorithm for calculating a mutual influence of the vertices in cognitive maps is that after increasing the values of each elements of matrix $W$ in the same value the vertices ranking does not change according to the degree of their influence. It means that the proposed algorithm does not violate the scale invariance after increasing of elements of the matrix $W$ in $\eta$. The general influence $Inf_{am}$ for each of the vertices is scaled up proportionally in $\eta$ after increasing every vertices in $\eta$. As the result of $\mu$-normalizing, $\eta$ enters into the expressions (3) and (4) as a multiplier.

For example, in the work [12] the cognitive map of the interplay of a model of sanitary condition (fig. 8) is considered.

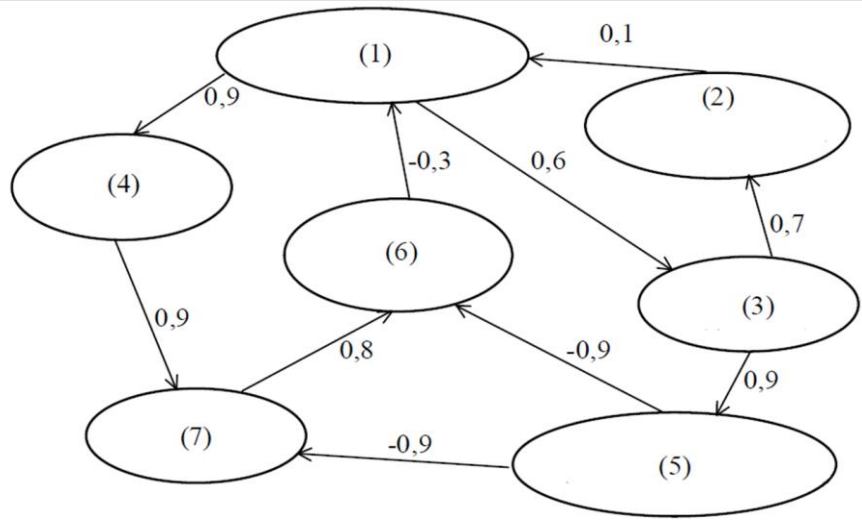

Fig. 8 The cognitive map of mutual influence of a sanitary condition model: 1 – Population of the city; 2 – Migration to the city; 3 – Modernization; 4 – Garbage dump; 5 – Sanitary condition; 6 – Morbidity; 7 – Bacteria

The adjacency matrix of the weighted directed graph presented on fig. 8 is

$$W = \begin{pmatrix} 0 & 0 & 0,6 & 0,9 & 0 & 0 & 0 \\ 0,1 & 0 & 0 & 0 & 0 & 0 & 0 \\ 0 & 0,7 & 0 & 0 & 0,9 & 0 & 0 \\ 0 & 0 & 0 & 0 & 0 & 0 & 0,9 \\ 0 & 0 & 0 & 0 & 0 & -0,9 & -0,9 \\ -0,3 & 0 & 0 & 0 & 0 & 0 & 0 \\ 0 & 0 & 0 & 0 & 0 & 0,8 & 0 \end{pmatrix} \quad (15)$$

The weighted directed graph defined by the adjacency matrix (15) is impulse stable for all simple impulse processes. All non-zero characteristic constants of the weighted directed graph are equal: $0,686$; $0,686$; $0,634$; $0,625$; $0,625$; $9,22e-17$.

According to the impulse method the general influence $Inf_p$ of each vertex and their ranking for the matrix (15) are presented in the tab. 7.

Table 7

| Vertex (No) | $Inf_p$ |
|---|---|
| 3 | 5,44 |
| 5 | 4,06 |
| 1 | 3,38 |
| 4 | 2,21 |
| 7 | 1,79 |
| 6 | 1,27 |
| 2 | 0,4 |

The influence matrix (15) is

$$Z = \begin{pmatrix} 0 & 0,515 & 0,6 & 0,9 & 0,662 & -0,008 & 0,684 \\ 0,1 & 0 & 0,12 & 0,179 & 0,055 & -0,002 & 0,0368 \\ 0,164 & 0,7 & 0 & 0,228 & 0,9 & -0,867 & -0,74 \\ -0,04 & -0,016 & -0,015 & 0 & -0,021 & 0,926 & 0,9 \\ 0,445 & 0,232 & 0,33 & 0,495 & 0 & -1,59 & -0,722 \\ -0,3 & -0,169 & -0,292 & -0,438 & -0,217 & 0 & -0,183 \\ -0,249 & -0,119 & -0,131 & -0,197 & -0,15 & 0,8 & 0 \end{pmatrix} \quad (16)$$

The general influence $Inf_{am}$ of each vertex and their ranking for the matrix (16) are presented in the tab. 8.

A vertices ranking (tab. 8) of the algorithm for calculating a mutual influence of the vertices implementation is similar by implication to a ranking as a result the impulse method use (tab. 7). The discrepancy of the ranking is a result of the algorithm for calculating a mutual influence of the vertices specification.

Table 8

| Vertex (No) | $Inf_{am}$ |
|---|---|
| 5 | 3,816 |
| 3 | 3,599 |
| 1 | 3,371 |
| 4 | 1,683 |
| 7 | 1,65 |
| 6 | 1,598 |
| 2 | 0,493 |

As a result of every element of an adjacency matrix $W$ increase by $\eta$, in case of the impulse method implementation, a scale invariance violation is observed, on the other hand, the algorithm for calculating a mutual influence of the vertices saves a vertices ranking in comparison to the ranking of the vertices of the matrix $W$.

$$W_2 = \begin{pmatrix} 0 & 0 & 1,2 & 1,8 & 0 & 0 & 0 \\ 0,2 & 0 & 0 & 0 & 0 & 0 & 0 \\ 0 & 1,4 & 0 & 0 & 1,8 & 0 & 0 \\ 0 & 0 & 0 & 0 & 0 & 0 & 1,8 \\ 0 & 0 & 0 & 0 & 0 & -1,8 & -1,8 \\ -0,6 & 0 & 0 & 0 & 0 & 0 & 0 \\ 0 & 0 & 0 & 0 & 0 & 1,6 & 0 \end{pmatrix} \quad (17)$$

For example, for the adjacency matrix (17) obtained as result of every elements of the adjacency matrix doubling (15), the vertices ranking in comparison to the vertices of matrix ranking (15) remains the same (tab. 9), on the other hand in this case the impulse method is not applicable because the directed graph defined by the adjacency matrix (17) is unstable within the meaning of impulse for some simple impulse process (an absolute values of all non-zero characteristic constants are equal: $1,37$; $1,37$; $1,27$; $1,25$; $1,25$; $1,84e-16$.

| Vertex (№) | $Inf_{am}$ for $W$ | $Inf_{am}$ for $W_2$ |
|---|---|---|
| 5 | 3,816 | 7,632 |
| 3 | 3,599 | 7,198 |
| 1 | 3,371 | 6,742 |
| 4 | 1,683 | 3,336 |
| 7 | 1,65 | 3,299 |
| 6 | 1,598 | 3,197 |
| 2 | 0,493 | 0,986 |

The results of the algorithm for calculating a mutual influence of the vertices implementation also show that the general influence $Inf_{am}$ for each of the vertices is scaled up proportionally in $\eta$ after increasing every vertex in $\eta$.

For example, for the adjacency matrix (15) after increasing it in 0,01; 0,1; 10 and 100 ($W_{0,01}, W_{0,1}, W_{10}$ та $W_{100}$) the general influence $Inf_{am}$ is scaled up proportionally in 0,01; 0,1; 10 and 100 (tab. 10).

Table 10

| Vertex (№) | $Inf_{am}$ for $W$ | $Inf_{am}$ for $W_{0,01}$ | $Inf_{am}$ for $W_{0,1}$ | $Inf_{am}$ for $W_{10}$ | $Inf_{am}$ for $W_{100}$ |
|---|---|---|---|---|---|
| 5 | 3,816 | 0,038 | 0,382 | 38,161 | 381,612 |
| 3 | 3,599 | 0,036 | 0,36 | 35,99 | 359,903 |
| 1 | 3,371 | 0,034 | 0,337 | 33,71 | 337,1 |
| 4 | 1,683 | 0,017 | 0,168 | 16,831 | 168,308 |
| 7 | 1,65 | 0,0165 | 0,165 | 16,496 | 164,96 |
| 6 | 1,598 | 0,016 | 0,16 | 15,982 | 159,827 |
| 2 | 0,493 | 0,005 | 0,049 | 4,928 | 49,276 |

**Conclusion**

Proposed algorithm for calculating mutual influence of the vertices in cognitive maps permits to overcome the disadvantages of the impulse method.

For example, for any finite number of vertices and for any values of the adjacency matrix describing the weighted directed graph the algorithm for calculating a mutual influence of the vertices gives a limited result $z_{i,j}$, unlike the impulse method. In the algorithm the initial values $z_{ij}^k(0)$ and $\tilde{z}_{ij}^k(0)$ influence the dependency $t$ of the values $z_{ij}^k(t)$ and $\tilde{z}_{ij}^k(t)$ (then $z_{ij}^k(0)=0$ and $\tilde{z}_{ij}^k(1)=0$). $z_{ij}^k(t)$ and $\tilde{z}_{ij}^k(t)$ according to the (2) and (3) do not depend on the initial impulse. Also after the values increase of each elements of matrix $W$ in $\eta$ the ranking of a vertices does not change according to the degree of their influence and the general influence $Inf_{sm}$ for each of the vertices is scaled up in $\eta$ proportionally. The algorithm for calculating a mutual influence of the vertices proposed in this paper has one disadvantage connected with a search of all simple ways between all pair of vertices of the cognitive map.

The search algorithm has an exponential complexity $O(e^{2n})$. However, sparse matrices implementation permits to employ the algorithm for calculating a mutual influence of the vertices for calculating the cognitive maps of large dimension analysis.